\documentstyle[11pt,newpasp,twoside,epsf]{article}
\def\HST{{\it HST}}

\def\kms{\ifmmode {\rm km\ s}^{-1} \else km s$^{-1}$\fi}
\def\ltsim{\raisebox{-.5ex}{$\;\stackrel{<}{\sim}\;$}}
\def\gtsim{\raisebox{-.5ex}{$\;\stackrel{>}{\sim}\;$}}
\def\civ{\ifmmode {\rm C}\,{\sc iv} \else C\,{\sc iv}\fi}
\begin{document}
\title{Properties of QSO-Intrinsic Narrow Ultraviolet Absorption}
\author{Rajib Ganguly}
\affil{Department of Astronomy \& Astrophysics, University Park, The Pennsylvania State University, State College, PA 16802}

\begin{abstract}
I present the current state of knowledge about narrow ($\ltsim$500
\kms) ultraviolet absorption that is intrinsic to QSOs. I consider
interpretations in the context of the accretion-disk/wind scenario of
QSOs.
\end{abstract}

\section{Introduction}

In the past four years, the study of truly intrinsic narrow absorption
has exploded both as a result of new insights about how to identify
these systems and because of the advent high resolution spectroscopy
with large ground-based telescopes. Before summarizing the work and
attempting to understand it in a coherent scenario, I start with a few
conventions regarding nomenclature. A narrow absorption line nominally
is one in which resonant UV doublets are well separated (Hamann \&
Ferland 1999). This implies widths less than about {500~\kms} in order
to resolve the {\civ$\lambda\lambda1548,1550$} doublet. I use the term
NALs to refer to narrow absorption lines (in general) and NALQSO to
refer to any QSO that has truly intrinsic narrow absorption (by
analogy with the term BALQSO). I also use the term ``associated''
absorption lines as signifying that the absorption lies within
{5000~\kms} of the QSO emission redshift. (Note that an associated
absorption line need not be intrinsic to the QSO.) Furthermore, a
strong system is one whose {\civ} rest-frame equivalent width is
larger than {1--2~\AA}. First, I will consider how we can identify
intrinsic systems -- that is, how we separate them from intervening
gas. I will then discuss separately what we know from studies at high
redshift and at low redshift. Finally, I will unify observed properties
in the context of the accretion-disk/wind scenario.

\section{Identifying Intrinsic NALs}

There are two basic methods which one uses to infer the properties of
intrinsic narrow absorbers. One can identify large populations of
absorbers in a sample and consider if their is a relationship with the
host QSO. Identifying populations of intrinsic absorbers typically
involves a demonstration that there is an excess of absorbing systems
over what is expected in a given redshift or velocity path. As a
consequence, it is now known which systems are actually intrinsic.

Alternatively, one can identify a specific absorber as intrinsic and
decipher the physical conditions of the gas. Identifying specific
intrinsic absorbers requires either multiple epochs of observation to
look for time variability and/or high resolution spectroscopy to show
that the absorbing gas only partly occults the QSO central engine.

Thus far, only fifteen QSOs have been shown conclusively to have
intrinsic NALs through time variability and/or partial coverage. Of
these, nine are radio-quiet, five are radio-loud, and only one is at
low redshift. Six have been shown to be time variable (TV) while ten
have been shown to exhibit the signature of partial coverage
(PC). These are listed in Table~1 along with the citation to the work
that showed the intrinsic origin.

\begin{table}
\caption{Known NALQSOs}
\begin{tabular}{llccl}
\tableline
Name             & {$z_{\rm em}$} & Radio & Method & Reference \\ \tableline
{Q~$0123+257$}   & {$2.358$}      & Loud  & PC    & Barlow \& Sargent (1997) \\
{Q~$0150-203$}   & {$2.139$}      & Loud  & PC, TV & Hamann et al. (1997a) \\
{PKS~$0424-131$} & {$2.166$}      & Loud  & PC     & Petitjean, Rauch, \& \\
                 &                &       &        & Carswell (1994) \\
{Q~$0449-134$}   & {$3.093$}      & Quiet & PC     & Barlow, Hamann, \& \\
                 &                &       &        & Sargent (1997) \\
{Q~$0450-132$}   & {$2.253$}      & Quiet & PC     & Ganguly et al. (1999) \\
{Q~$0835+580$}   & {$1.534$}      & Loud  & TV     & Aldcroft, Bechtold, \& \\
                 &                &       &        & Foltz (1997) \\
{Q~$0935+417$}   & {$1.980$}      & Quiet & TV     & Hamann et al. (1997b) \\
{PKS~$1157+014$} & {$1.986$}      & Loud  & TV     & Aldcroft, Bechtold, \& \\
                 &                &       &        & Foltz (1997) \\
{PG~$1222+228$}  & {$2.038$}      & Quiet & PC     & Ganguly et al. (1999) \\
{PG~$1329+412$}  & {$1.930$}      & Quiet & PC     & Ganguly et al. (1999) \\
{HS~$1700+6416$} & {$2.722$}      & Quiet & PC, TV & Barlow, Hamann, \& \\
                 &                &       &        & Sargent (1997) \\
{Q~$2116-358$}   & {$2.341$}      & Quiet & PC     & Wampler, Bergeron, \& \\
                 &                &       &        & Petitjean (1993) \\
{QSO~J$2233-606$} & {$2.238$} & ...& PC   & Petitjean \& Srianand (1999) \\
{MRC~$2251-178$} & {$0.066$}      & Loud  & TV     & Ganguly, Charlton, \& \\
                 &                &       &        & Eracleous (2001b) \\
{Q~$2343+125$}   & {$2.515$}      & Quiet & PC, TV & Hamann et al. (1997c) \\
\tableline
\tableline
\end{tabular}
\end{table}

\section{Intrinsic NALs at High Redshift}

At high redshift, since the rest-frame ultraviolet transitions are
shifted into the optical, we can take advantage of optical
spectroscopy with large ground-based telescopes to identify specific
absorbing systems as intrinsic. Unfortunately, since the QSOs are at
higher redshift, it is generally more difficult to obtain detailed
multiwavelength information about the QSOs themselves.

Strong systems seem to prefer optically-faint (OF), radio-loud QSOs
with steep radio spectra (Foltz et~al. 1986; Anderson et~al. 1987;
M{\o}ller \& Jakobsen 1987; Foltz et~al. 1988). In addition, the
equivalent width of strong systems seems to correlate with
orientation, with stronger systems existing in more radio-lobe
dominated (that is, edge-on) QSOs (Barthel, Tytler, \& Vestergaard
1997; Baker et~al., these proceedings). A recent study of QSO
absorption systems down to a {0.15~\AA} limiting equivalent width also
found that intrinsic systems can appear at very large ``ejection''
velocities (Richards et~al. 1999, Richards 2001). This happens more so
in radio-quiet QSOs than radio-loud QSOs and more so in flat-spectrum,
radio-loud (FSRL) QSOs than steep-spectrum, radio-loud (SSRL) QSOs.

\section{Intrinsic NALs at Low Redshift}

At low redshift, we have the advantage of knowing very well the QSO
properties. However, it is harder to identify intrinsic systems since
we must rely on smaller space-based telescopes ({\HST} and {\it
FUSE}). High-resolution spectra can be obtained for only the
brightest targets.

The first remarkable property of low-redshift NALQSOs is that none
host strong systems (Ganguly et~al. 2001a). As a result, the
correlations with QSO properties seen at high redshift, which were
driven by strong absorption, are largely absent. [Strong systems seem
to exist in compact, steep-spectrum (CSS) radio-loud QSOs down to
{$z_{\mathrm{em}}\sim0.7$} (Baker et~al., these proceedings).] The
equivalent widths of weak systems do not correlate with any single
QSOs property but their velocity distribution seems to peak at the
same velocity as the broad emission lines. This seems to indicate a
relationship between the line-of-sight velocity of the absorbers and
the velocity of maximum emissivity of the broad line region. A
multivariate analysis of associated absorption indicates a combination
of QSO properties that seem to prohibit the detection of associated
NAL gas. Associated NALs are absent in FSRL QSOs that have mediocre
{\civ} emission FWHM ($\ltsim6000$~\kms), but are present in a finite
fraction of FSRL QSOs with large {\civ} FWHM.

\begin{table}
\caption{Properties of QSO-Intrinsic NALs}
\begin{tabular}{rlrl}
\tableline
\multicolumn{2}{c}{$z_{\rm em} \gtsim 1$}  & \multicolumn{2}{c}{$z_{\rm em} \ltsim 1$} \\ \tableline
\multicolumn{4}{c}{Strong systems} \\ \tableline
{$\bullet$} & Prefer OF SSRL QSOs               & {$\bullet$} & Largely absent @ {$z \ltsim 0.7$} \\
{$\bullet$} & EW correlated with orientation    &             & but exist in CSS RL QSOs @        \\
{$\bullet$} & Velocity distribution peaks with  &             & {$z \gtsim 0.7$}                  \\
            & BEL, not {$z_{\rm sys}$}          &             &                                   \\
{$\bullet$} & Exist at high {$v_{\rm ej}$}      &             &                                   \\
\tableline
\multicolumn{4}{c}{Weak systems} \\ \tableline
{$\bullet$} & No preference for RL              & {$\bullet$} & Absent in FSRL QSOs with          \\
{$\bullet$} & Excess of high {$v_{\rm ej}$} NALs in &     & {\civ} BEL FWHM \ltsim 6000 \kms  \\
            & RQs compared to RLs, and          & {$\bullet$} & Velocity distribution peaks       \\
            & FSRLs compared to SSRLs           &             & with BEL, not systemic velocity   \\
            &                                   & {$\bullet$} & Enhanced probability of           \\
            &                                   &             & NALs in BALQSOs                   \\
\tableline
\tableline
\end{tabular}
\end{table}

\section{Putting It All Together}

These properties are summarized in Table 2, which is broken up
according to both redshift and absorption strength. If we postulate
that a unified model exists, there are three basic conclusions to draw
from the table. First, there has been evolution such that strong
systems are largely absent at low redshift. Second, the properties of
weak absorbers at both, high and low redshift complement each other so
that no evolution in their population is required. That is, the
properties at both high and low redshift can be considered together as
governing an unevolving population of absorbers. Similarly, the
properties of high redshift strong systems and low redshift weak
systems do not contradict each other and can be considered
simultaneously.

\begin{figure}
\plotone{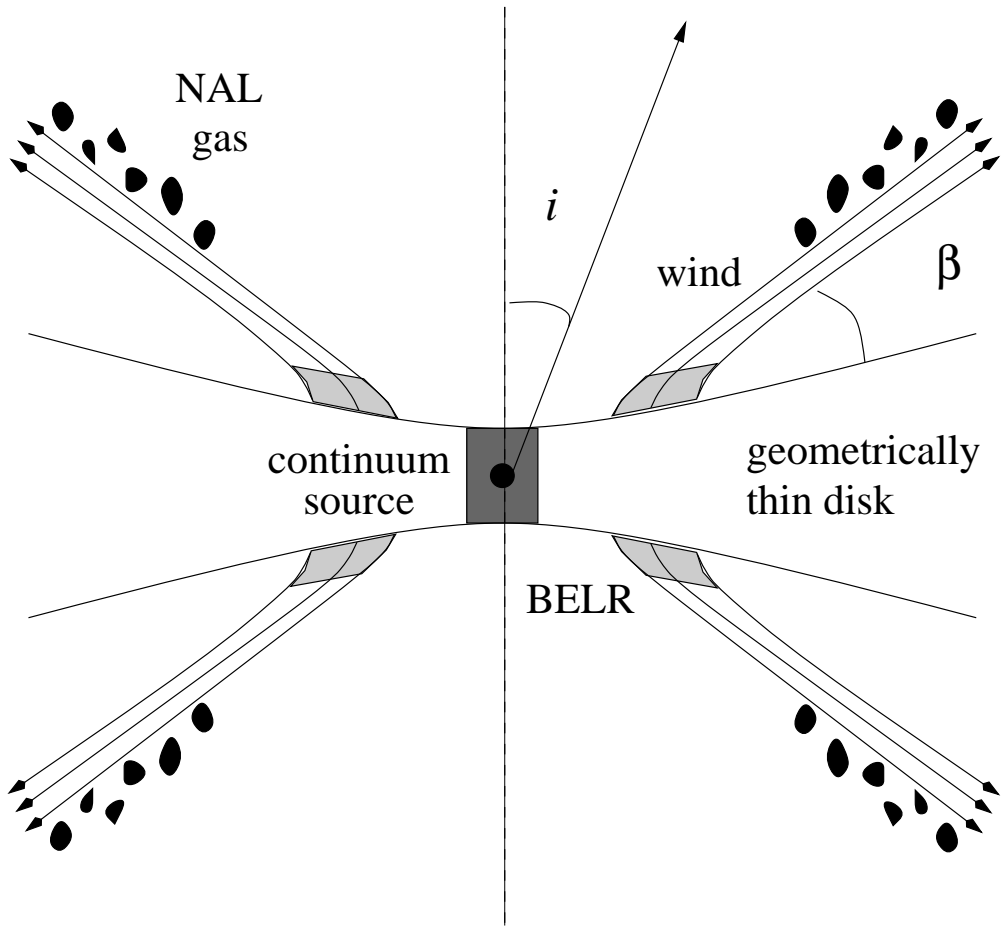}
\caption{Disk-wind model for QSOs from Ganguly et al. (2001a). The
inclination angle, {\it i}, and the wind opening angle, {$\beta$}, are
shown.}
\label{fig:diskwind}
\end{figure}

We can understand these properties in the context of the disk-wind
scenario. A cartoon of this scenario, from Ganguly et al. (2001a), is
shown in Fig. 1. The scenario, a modification of the Murray et
al. (1995) model originally employed to explain broad absorption
lines, is that of a radiatively driven, outflowing wind in which
clumps of gas hug the wind at ``large'' distances from the black
hole. Hydrodynamic simulations by Proga, Stone, \& Kallman (2000) show
that such clumps do arise from Kelvin-Helmholtz shearing
instabilities.

This scenario has, essentially three fundamental parameters: the black
hole mass, the mass fueling rate, and the inclination with respect to
the observer. The black hole mass and mass fueling rate can also be
translated into a wind opening angle and wind density. A given black
hole mass implies a maximum mass accretion rate (that is, the
Eddington rate). For a larger mass fueling rate, this implies a larger
wind density. Moreover, the luminosity of the QSO will be larger and
thus the acceleration of the wind will be even more dominated by its
radial component. This will result in a smaller wind opening angle.

If we hypothesize that the wind in radio-loud QSOs is less dense than
in radio-quiet QSOs, then strong NALs can be thought of as the
analogues of BAL, where the wind itself is the source of
absorption. Reversing the above reasoning, a sparser wind implies a
less luminous QSO, explaining the preference for ``optically faint''
QSOs. The projected velocity dispersion of the wind along sightlines
is small (i.e. narrow). Likewise, the optical depth of the wind is
largest when the QSO is viewed at higher inclination angle (similar to
BALQSOs).

The evolution of strong systems can be viewed as a change in the mass
outflow rate (or wind density). This change can result either from a
decrease in the mass fueling rate or an increase in the mass accretion
rate. Either case seems natural since (1) there is only a finite
amount of gas to fuel the engine and (2) over the duty cycle of the
QSOs life, the accretion process will increase the black hole mass,
and therefore the maximum allowed accretion rate.

The absence of weak associated systems in FSRL QSOs w/ average CIV
FWHM can be seen as mostly an inclination effect. If the population of
weak absorbers is due the clumps produced by the shear, then NALs will
only be detected when the line of sight intercept these clumps. The
flat-spectrum radio-loud QSOs in the Ganguly et al. (2001) sample were
also strongly radio core dominated. So, there is little doubt that the
QSOs are viewed at small inclination angles (i.e. face-on
geometries). In addition, the width of the {\civ} emission line
implies that the velocity dispersion of the wind along the line of
sight is not large. Thus, the wind opening angle is ``small.'' In this
case, neither photons from the compact continuum, nor photons from the
broad emission line region intercept the weak NAL clouds.

\acknowledgements
This work was funded by NASA through grants NAG 5-6399,
HST-GO-08681.01-A, and through an archival award from the Space
Telescope Science Institute (STSI AR-08763.01-A), which is operated by
AURA, Inc., under NASA contract NAS 5-26555. Travel expenses to the
meeting were provided by the Zaccheus Daniel Foundation.
R.G. acknowledges Ray Weymann for a thoughtful meeting summary, and a
stimulating discussion after the meeting.

\end{document}